# PRODUCCIÓN DE ENTROPÍA EN CIRCUITOS ELÉCTRICOS SENCILLOS

# ENTROPY PRODUCTION BY SIMPLE ELECTRICAL CIRCUITS

**E. N. Miranda[1,2,3*], S. Nikolskaia[1]**

[1] Facultad de Ingeniería, Universidad de Mendoza, 5500 Mendoza, Argentina
[2] Departamento de Física, Universidad Nacional de San Luis, 5700 San Luis, Argentina
[3] Área de Ciencias Exactas, CRICYT - CONICET, 5500 – Mendoza, Argentina

Se analiza la producción de entropía en circuitos R, RC y RL. Se muestra que la producción es mínima en el estado estacionario de acuerdo a un conocido teorema de Prigogine. Se refuta así la afirmación hecha recientemente por Zupanovic, Juretic y Botric (Physical Review **E 70**, 056108 (2004) quienes sostienen que la entropía es maximizada por los circuitos eléctricos simples.

Palabras clave: producción de entropía, teorema de Prigogine, termodinámica de no equilibrio

The entropy production in simple electrical circuits (R, RC, RL) is analyzed. It comes out that the entropy production is minimal, in agreement with a well known theorem due to Prigogine. In this way, it is wrong a recent result by Zupanovic, Juretic and Botric (Physica Review **E 70**, 056198 ) who claimed that the entropy production in simple electrical circuits is a maximum.

Key works: entropy production, Prigogine´s theorem, non-equilibrium thermodynamics

La termodinámica de sistemas fuera de equilibrio es tema controversial, no habiéndose logrado aún un consenso unánime en la comunidad científica. En este sentido aún hoy hay disputas acerca de la generación de entropía por sistemas tan sencillos como un circuito eléctrico[1], [2], [3]. Existe un teorema debido a Prigogine [1] que establece que la producción de entropía en un sistema estacionario es mínima. En tal sentido, cabría esperar que la producción de entropía por un circuito eléctrico constituido por fuentes y resistencias fuera mínima. Sin embargo, recientemente esta afirmación ha sido cuestionada [2]. Los autores de la Ref. 2, Zupanovic, Juretic y Botric a quienes llamamos ZJB de ahora en más, sostienen que la producción de entropía en tales circuitos es máxima. Ellos realizan un análisis matemático de problema que aparenta gran solidez, introduciendo una función *F* que estaría vinculada con la producción de entropía y muestran que *F* se maximiza en el estado estacionario de un circuito constituido por fuentes y resistencias. En forma independiente, Browne [3] resuelve explícitamente varios circuitos sencillos de tipo RC, RL y RLC en los que el teorema no se cumpliría en los estados transitorios. En este trabajo se muestra que en los más elementales de tales circuitos la producción de entropía se minimiza. De esta manera se refuta el análisis realizado por ZJB y se refuerza la convicción en que el teorema de Prigogine es válido.

---
[*] E-mail: emiranda@lab.cricyt.edu.ar

Se comienza analizando un circuito elemental formado por una fuente de tensión $V$ y dos resistencias conectadas en paralelo $R_1$ y $R_2$ tal como se muestra en la Figura 1.

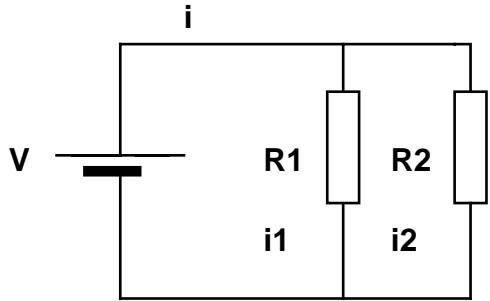

**Figura 1:** *Un circuito simple formado por dos resistencias en paralelo con una fuente. Se muestra explícitamente que la producción de entropía es mínima.*

La producción de entropía es la debida a la producción de calor en las resistencias. Si la temperatura ambiente es $T$, entonces la producción de entropía por unidad de tiempo $\pi$ será:

$$\pi = \frac{i_1^2 R_1 + i_2^2 R_2}{T} \quad (1)$$

Teniendo en cuenta la conservación de la carga, se puede escribir:

$$\pi = \frac{i_1^2 R_1 + (i - i_1)^2 R_2}{T} \quad (2)$$

Si se supone que la producción de entropía es un mínimo, entonces:

$$\frac{\partial \pi}{\partial i_1} = 0$$

$$i_1 = \frac{R_2}{R_1 + R_2} i = \frac{V_0}{R_1} \quad (3)$$

$$i_2 = \frac{R_2}{R_1 + R_2} i = \frac{V_0}{R_2}$$

Estos valores de $i_1$ e $i_2$ son exactamente los que se obtienen por métodos usuales. Y no es difícil demostrar que minimizan la entropía. Calculando la derivada segunda, resulta:

$$\frac{\partial^2 \pi}{\partial i_1^2} = R_1 + R_2 > 0 \quad (4)$$

El análisis de ZJB [2] aparece como muy convincente, sin embargo el ejemplo elemental mostrado muestra fuera de cualquier duda que un circuito con una fuente de voltaje y resistencias en paralelo minimiza la producción de entropía. ZJB introducen una función $F$ y ellos muestran que se maximiza. Sin embargo, es claro que la maximización de $F$ no significa la maximización de $\pi$.

Si ahora consideramos una fuente de tensión conectada a una resistencia y a un capacitor $C$, resulta trivial que en el estado estacionario la producción de entropía es mínima. En efecto, en estado estacionario la corriente que circula por el circuito es nula y por lo tanto el calor producido es estrictamente cero. Podemos considerar que se trata de un capacitor real, caracterizado tanto por su capacidad $C$ como por una resistencia de pérdida $R_C$ tal como se muestra en la Figura 2.

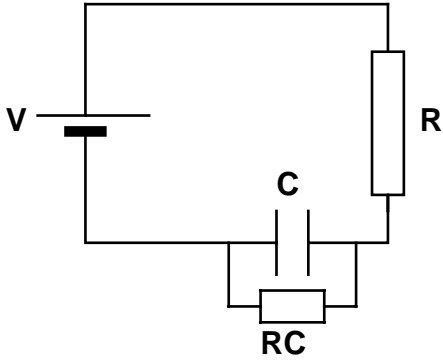

**Figura 2:** *Una fuente de tensión V conectada en serie con una resistencia R y un capacitor real, caracterizado por una capacidad C y una resistencia de pérdida Rc.*

En estado estacionario, simplemente se tienen dos resistencias conectadas en serie por las que circula una corriente *i*. La producción de entropía por unidad de tiempo es:

$$\pi = \frac{1}{T}\left(\frac{V_R^2}{R} + \frac{V_C^2}{R_C}\right) \quad (5)$$

Siendo $V_R$ y $V_C$ la caída de potencial en la resistencia y en el capacitor respectivamente. Por conservación de la energía, tiene que ser:

$$V = V_R + V_C \quad (6)$$

En consecuencia, suponiendo que la entropía se minimiza:

$$\frac{d\pi}{dV_R} = \frac{1}{2T}\left(\frac{V_R}{R} - \frac{V - V_R}{R_C}\right) = 0 \quad (7)$$

Despejando resulta:

$$V_R = \frac{R}{R + R_C}V$$
$$V_C = \frac{R_C}{R + R_C}V \quad (8)$$

Y estos son los valores correctos. Para verificar que se trata de un mínimo, la derivada segunda arroja el siguiente resultado:

$$\frac{d^2\pi}{dV_R^2} = \frac{1}{2T}\left(\frac{1}{R} + \frac{1}{R_C}\right) > 0 \quad (9)$$

En síntesis, se ha demostrado que un circuito formado por un capacitor real en serie con una resistencia minimiza la entropía.

Queda por analizar el caso de una resistencia conectada en serie con una inductancia real, caracterizada por una autoinducción *L* y una resistencia interna $R_L$ tal como se muestra en la Figura 3.

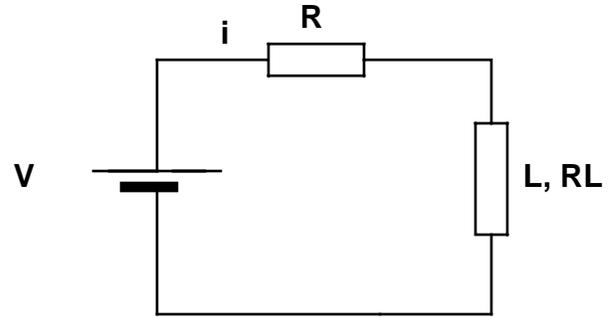

**Figura 3:** *Una fuente de tensión V conectada en serie con una resistencia R y una inductancia real, caracterizado por una autoinducción L y una resistencia interna $R_L$. Se muestra en el texto que la entropía es minimizada en el estado estacionario.*

En estado estacionario la autoinducción no afecta el valor de la intensidad *i*. Por lo tanto el circuito se reduce a dos resistencias conectadas en serie que como se mostró en el caso del capacitor, minimiza la producción de entropía.

En síntesis, en esta trabajo se ha mostrado que circuitos sencillos del tipo RLC, conectados a una fuente de tensión continua, minimizan la producción de calor y por ende de entropía cuando los valores de la intensidad y las caídas de potencial toman los valores obtenidos por aplicación de las leyes de Kirchoff. En tal sentido, parece posible afirmar que la refutación presentada en la Ref. 2 es errónea. Respecto a lo sostenido en la Ref. 3 sobre algunos circuitos en particular y sus transitorios parece ser correcta tal como se reporta en otro lugar [4]. En todo caso, eso no contradice el resultado de Prigogine ya que el mismo se refiere a estados estacionarios.

En el futuro se estudiará el comportamiento de circuitos eléctricos de corriente alterna y con elementos no-lineales, tanto en estado estacionario como sus transitorios.